\documentclass[twocolumn,prbrc]{revtex4}
 
\usepackage{graphicx}
\usepackage{dcolumn}
\usepackage{amsmath}
\begin{document}

\preprint{Submitted to {\it Phys. Rev. B RC}}


\title{Propagating Coherent Acoustic Phonon Wavepackets in
InMnAs/GaSb}

\author{J. Wang, Y. Hashimoto,\footnote{
Permanent address: Graduate School of Science
and Technology, Chiba University, Chiba, Japan} J. Kono}
\thanks{Author to whom correspondence should be addressed}
\email{kono@rice.edu}
\affiliation{Department of Electrical and Computer Engineering,
Rice Quantum Institute, and Center for Nanoscale Science and Technology,
Rice University, Houston, Texas 77005}

\author{A. Oiwa}
\thanks{Present address: Department of Applied Physics, University of
Tokyo, 7-3-1 Hongo, Bunkyo-ku, Tokyo 113-8656, JAPAN}
\affiliation{PRESTO, Japan Science and Technology Agency, 4-1-8 Honcho, Kawaguchi 332-0012, Japan}

\author{H. Munekata}
\affiliation{Imaging Science and Engineering Laboratory,
Tokyo Institute of Technology, 4259 Nagatsuta, Yokohama 226-8503, Japan}

\author{G. D. Sanders, C. J. Stanton}
\affiliation{Department of Physics, University of Florida,
Gainesville, Florida 32611}

\date{\today}


\begin{abstract}
We observe pronounced oscillations in the differential
reflectivity of a ferromagnetic InMnAs/GaSb heterostructure
using two-color pump-probe spectroscopy. Although originally
thought to be associated with the ferromagnetism, our studies
show that the oscillations instead result from changes in the
position and frequency-dependent dielectric function due to the
generation of coherent acoustic phonons in the ferromagnetic
InMnAs layer and their subsequent propagation into the GaSb.
Our theory accurately predicts the experimentally measured
oscillation period and decay time as a function of probe
wavelength.

\end{abstract}
\maketitle



Recently, there has been much interest in (III,Mn)V dilute
magnetic semiconductors (DMS) with carrier-mediated
ferromagnetism, a promising system for the realization of future
semiconductor spintronic devices capable of performing
information processing, data storage, and communication
functions simultaneously \cite{Zutic04.323,Ohno98.951,
Wolf01.1488}. InMnAs is the prototypical III-V DMS, being the
first DMS to exhibit ferromagnetism, and the first DMS in which
cyclotron resonance was observed, evidence that at least some of
the carriers are itinerant
\cite{ZudovetAl02PRBRC,SandersetAl03PRB,MatsudaetAl04PRB}.

Despite the fact that time-domain studies
of the dynamic aspects of (III,Mn)Vs are more informative than
static magnetization or electrical transport measurements,
relatively few time-dependent studies have been attempted. In
this paper, we report on our time-dependent femtosecond
transient reflectivity  measurements on an InMnAs/GaSb
heterostructure using two-color pump-probe spectroscopy.  In
addition to changes in the reflectivity associated with
utlrashort carrier lifetimes ($\sim$ 2 ps) and multi-level
carrier decay dynamics which we attribute to a large density of
bound states and a high concentration of Mn $p$-type dopants, we
observe pronounced oscillations in the differential
reflectivity signal. While originally the oscillations were
believed to originate from the ferromagnetism (since they were
not observed in similiar systems without Mn doping), a
systematic study shows that they instead result from the
generation of coherent phonons in the InMnAs layer which then
propagate into the GaSb. The period and the characteristic decay
time of the transient reflectivity oscillations are consistent
with a propagating coherent acoustic phonon wavepacket
model \cite{Liu03.0310654}.

While femtosecond spectroscopy has previously been used for
generating and detecting coherent phonons in many materials
(e.g, optical phonons in bulk semiconductors, metals (Bi, Sb)
and superconductors; acoustic phonons in InGaN/GaN-based
semiconductor heterostructures \cite{Cho90.764, Bartels99.1044,
Sun00.179, Yahng02.4723, Liu03.0310654, Cheng91.1923,
Chwalek91.980}), only recently has there been interest in the
application of this technique to probing coherent
excitations (e.g., phonons and magnons) in strongly-correlated
electronic systems
\cite{Lim03.4800, Arita02.127202}. These innovative studies
further demonstrate the potential of transient reflectivity
spectroscopy to unravel the subtle interplay between lattice and
spin excitations.
%
\begin{figure}[tbp]
\begin{center}
\includegraphics[scale=0.7]{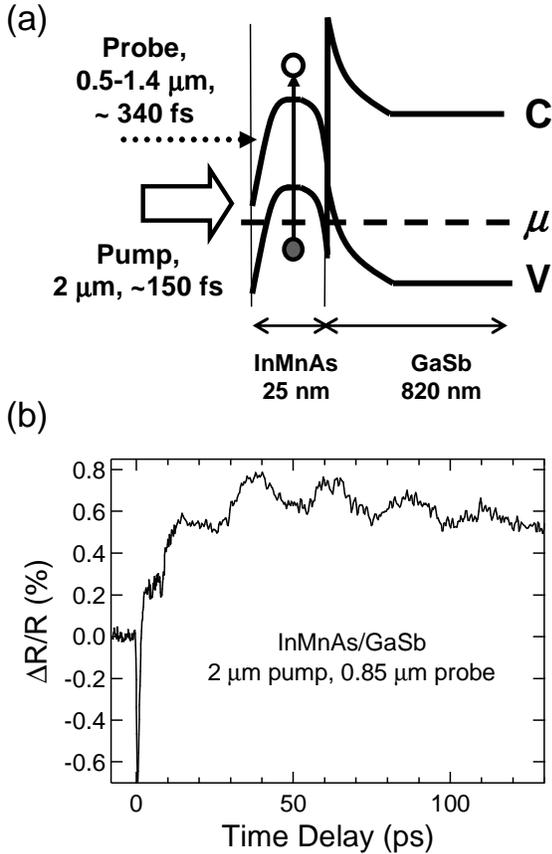}
\caption{
(a) Schematic diagram of two-color pump-probe differential
reflectivity experiment in an InMnAs/GaSb heterostructure. The
pump photoexcites carriers only in the InMnAs layer and the
transient differential reflectivity is probed at higher
energies. (b) Typical time-resolved differential
reflectivity data, showing an initial drop followed by
oscillations in reflectivity.}  \label{fig1} \end{center}
\end{figure}
 
The main Mn-doped sample studied was a $p$-InMnAs/GaSb
heterostructure grown by low-tempeature molecular beam
epitaxy (LT-MBE). The sample [shown schematically in
Fig.~\ref{fig1}(a)] consisted of a 25-nm $p$-type InMnAs
layer and an 820-nm GaSb buffer layer grown on top of a
(100) GaAs substrate. The hole density and mobility in the InMnAs
layer, as estimated by Hall measurements, were 1.1 $\times$
10$^{19}$ cm$^{-3}$ and 323 cm$^2$/Vs, respectively. Detailed
growth conditions and sample information can be found in
Ref.~\onlinecite{Slupinski02.1326}.

In order to separate various nonlinear effects and extract
information on the dynamical response of the InMnAs/GaSb
heterostructure, we used a two-color, selective pumping scheme
\cite{Wang03.373} shown schematically in Fig.~\ref{fig1}(a). A
140 fs, midinfrared (MIR) 2 $\mu$m pump beam from an optical
parametric amplifier (OPA) was used to create carriers in InMnAs
with an excess kinetic energy of $\sim$0.2 eV. Pumping at this
photon energy allowed us to selectively photoexcite carriers in
the InMnAs layer but not in the GaSb buffer layer or GaAs
substrate. A near-infrared (NIR) (775 nm) probe beam from a
chirped pulse amplifier (Model CPA-2010, Clark-MXR, Inc.)
allowed us to probe energies far above the quasi-Fermi level of
the optically excited carriers.
 
Typical time-dependent differential reflectivity data are shown
in Fig.~\ref{fig1}(b). After photoexcitation, the reflectivity
shows a sharp drop followed by a rapid rise and sign change in a
time less than $\sim$2 ps.  At longer times (several hundred ps)
periodic oscillations are observed with a period of $\sim$23 ps
superimposed on a very slow decay. The initial sharp drop in
reflectivity is believed to result from free carrier Drude
absorption by the hot photogenerated carriers.  The carriers
relax back to quasi-equilibrium distributions through the
emission of confined LO phonons and the ultrafast trapping of
electrons (by As$_{Ga}$ antisite defects) and holes (by Ga
vacancies) by the mid-gap states introduced by LT-MBE growth.
This alters the dielectric function of the heterostructure
through changes in the electron and hole distribution functions
and gives rise to the sharp increase and sign change in the
differential reflectivity seen in Fig.~\ref{fig1}(b). Over much
longer times, the quasi-equilibrium electrons and holes recombine
across the gap causing the differential reflectivity to return to
zero. The beginning of this slow decay can be seen in
Fig.~\ref{fig1}(b).
%
\begin{figure}[tbp]
\begin{center}
\includegraphics[scale=0.8]{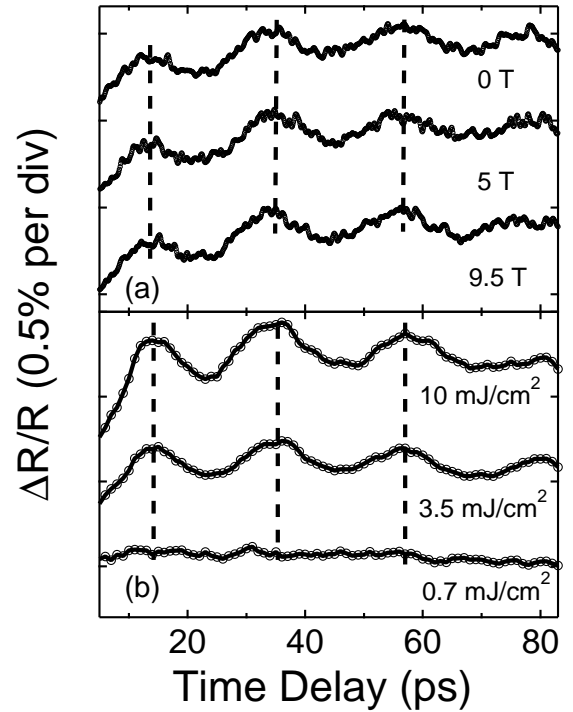}
\caption{
Two-color pump-probe differential reflectivity oscillations in an
InMnAs/GaSb heterostructure. The dependence of the oscillations on
magnetic field is shown in (a) while the dependence on pump power
is shown in (b). The oscillation period does not change
as shown by the vertical dashed lines.
}
\label{fig2}
\end{center}
\end{figure}
%

Initially, the oscillations were thought to originate from the Mn
ions and possibly the ferromagnetism since they were not observed
in similiar samples without the Mn.  However, when we applied an
external magnetic field of 0, 5, and 9.5 Tesla, the
oscillation period did not change, as shown in
Fig.~\ref{fig2}(a).  This rules out magnons or
quantum beats between Landau levels as a source of the
oscillations.

Next, we varied the pump fluence as shown in Fig.~\ref{fig2}(b).
For pump fluences of 0.7, 3.5, and 10 mJ/cm$^2$,  the intensity
of the oscillation changed but the frequency did not.  Varying
the pump fluence increases the photoexcited carrier density.
Since the plasma frequency increases with the total carrier
concentration, we would expect the oscillation period to increase
with increasing pump fluence if it were related to plasmons.
Since the oscillation period is independent of pump fluence, we
rule out plasmons as the cause of the oscillations.

A final possibility is that coherent phonons could be excited in
the InMnAs epilayer, since they have previously been observed in
InGaN/GaN epilayers \cite{Yahng02.4723,Liu03.0310654}, though
InGaN/GaN has a wurtzite structure and as a result, strain can
lead to large built-in piezoelectric fields (unlike InMnAs which
has zinc-blende structure).  The theory for the generation and
propagation of coherent acoustic phonons in InGaN/GaN
nanostructures having the wurzite structure has been described
in Ref.~\onlinecite{Sanders01.235316} (see the erratum in
Ref.~\onlinecite{Sanders02.079903}).  Reference
\onlinecite{Chern04.339} provides a good review of both
experimental and theoretical aspects of coherent acoustic phonon
generation in piezoelectric nitride-based semiconductor
heterostructures.  Generation of coherent acoustic phonons comes
about from the electron-phonon interaction
\cite{Kuznetsov94.3243,Kuznetsov95.7555,Kuznetsov01.353}. In
nitride-based heterostructures such as GaN/InGaN large amplitude
coherent acoustic phonon wavepackets can be generated through the
piezoelectric and deformation potential electron-phonon
interactions with the piezoelectric interaction being dominant.

In contrast to the nitrides, the InMnAs/GaSb heterostructure has
zinc-blende structure.  Consequently, coherent acoustic phonons
cannot be generated (for [100]-grown samples) through the
{\em piezoelectric} electron-phonon interaction.  They can
however be generated through the {\em deformation potential}
electron-phonon interaction, but their amplitudes are weak.
However, we can detect the coherent phonon oscillations in the
differential reflectivity through a fortuitous circumstance. The
wavelength-dependent GaSb dielectric function is very sensitive
to small changes in strain in the vicinity of the $E_1$
(L-valley) transition centered at approximately 600 nm, which is
close to our probe region.

We developed a theoretical model for the transient differential
reflectivity in our InMnAs/GaSb structure based on a Boltzmann
equation formalism. The photoexcited carriers in the
ferromagnetic quantum well are assumed to be completely confined
and electronic states near the band edge are treated in an
eight-band effective mass model including conduction
electrons, heavy holes, light holes, and split-off holes. The
effect of Mn impurity spins on the itinerant carriers is also
included. Photogeneration of carriers in the quantum well is
treated using Fermi's Golden Rule and carrier scattering by
confined LO phonons in the quantum well is accounted for. From
the time-dependent carrier densities, the generation and
propagation of coherent acoustic phonons can be modelled by
solving a loaded string equation as described in
Ref.~\onlinecite{Sanders01.235316}. Changes in the position- and
frequency-dependent dielectric function due to coherent acoustic
phonons are computed and the time-dependent reflectivity at the
probe wavelength is obtained by globally solving Maxwell's
equations in the entire structure.

%
\begin{figure}[tbp]
\begin{center}
\includegraphics[scale=0.8]{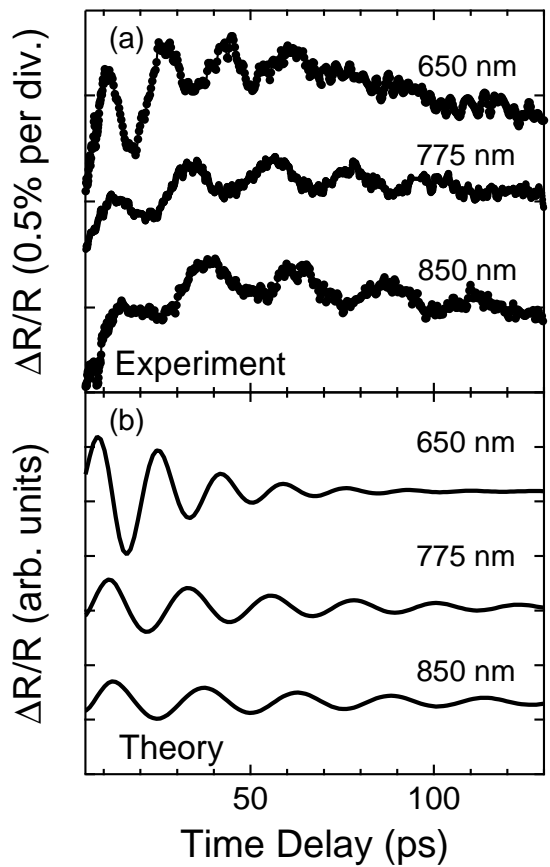}
\caption{
Experimental (a) and theoretical (b) coherent phonon
differential reflectivity oscillations for probe wavelengths of
650, 775 and 850 nm.
}
\label{fig3}
\end{center}
\end{figure}

In Fig.~\ref{fig3} the coherent phonon differential reflectivity
oscillations are shown as a function of time delay for probe
wavelengths of 650, 775, and 850 nm. The theoretical differential
reflectivity curves in Fig.~\ref{fig3}(b) agree well with the
experimentally measured differential reflectivity seen in
Fig.~\ref{fig3}(a) after subtraction of the transient
background signal. As the probe wavelength approaches the GaSb
$E_1$ transition near 600 nm, the amplitude of the differential
reflectivity oscillations increase due to the enhanced
sensitivity of the GaSb dielectric function to the localized
propagating strain pulse.

The damping of the differential reflectivity oscillations seen in
Fig.~\ref{fig3} is due to the finite penetration length of our
probe inside the GaSb layer. The periodic oscillations in the
differential reflectivity can be modelled by
\begin{equation}
\nonumber
\frac{\triangle R}{R}(t)
\sim \alpha \ \cos \left( {4\pi C_{s}n\over \lambda}\ t-\Phi \right)
e^{-t/\cal{T}},
\end{equation}
where $\Phi$ and ${\cal{T}}$ are the phase and decay time,
$\lambda=2\pi \ c/\omega$ is the probe wavelength, $C_s$ is the
acoustic sound speed, and $n$ is the refractive index at the
probe wavelength. From fits to the data, the decay time for a
probe wavelength of 650 nm is found to be $\sim$ 45 ps. The
decay time is approximately ${\cal{T}}=\lambda /(4\pi \ C_{s} \
k)$, where $k$ is the extinction coefficient at the probe
wavelength. A numerical estimate yields a value of 41 ps for the
decay time in reasonable agreement with experiment seen in
Fig.~\ref{fig3}.

%
\begin{figure}
\begin{center}
\includegraphics [scale=0.8]{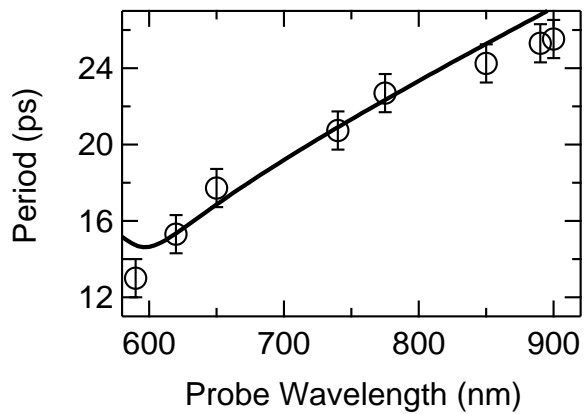}
\caption{
Experimental coherent phonon differential reflectivity
oscillation period vs. probe wavelength (open circles).
The solid line shows the theoretical estimates. }
\label{fig4}
\end{center}
\end{figure}

The differential reflectivity oscillation period increases with
wavelength from 650 to 850 nm. The period of the reflectivity
oscillations can be easily understood. The propagating strain
pulse gives rise to a perturbation in the GaSb dielectric
function which propagates at the acoustic sound speed. The
sample thus acts as a Fabry-Perot interferometer and a simple
geometrical optics argument shows that the period for the
reflectivity oscillations due to the propagating coherent
acoustic phonon wavepacket is approximately \cite{Yahng02.4723}
$T=\lambda/( 2 \ C_s \ n(\lambda) )$ where $C_s = 3.97 \times
10^{5}$ cm/s is the LA sound speed in the GaSb barrier and
$n(\lambda)$ is the wavelength dependent refractive index. At
the wavelength of the probe (850 nm), $n(\lambda) \approx 4.24$
and we calculate the period of the differential reflectivity
oscillations to be $T = 25.2$ ps, which is in close agreement
with the observed oscillation period of 24 ps seen in
Fig.~\ref{fig1}~(b). In Fig.~\ref{fig4} we have plotted the
experimentally measured coherent phonon differential
reflectivity oscillation periods as a function of probe
wavelength as open circles. The solid line shows the oscillation
period vs. probe wavelength estimated using the simple
geometrical optics argument and taking the wavelength dependence
of the refractive index in GaSb into account. The excellent
agreement between theory and experiment clearly demonstrates
that the reflectivity oscillations are the result of propagating
coherent acoustic phonons in the GaSb barrier.

In summary, we have performed time-dependent
two-color differential reflectivity measurments on a
ferromagnetic InMnAs/GaSb heterostructure.  In addition to
observing changes in the reflectivity due to ultrashort carrier
dynamics, we have observed reflectivity oscillations resulting
from coherent phonon generation in the InMnAs layer.  The
propagation of these coherent, localized ``strain pulses''
into and through the GaSb buffer layer results in a position and
frequency-dependent dielectric function.  Our
theoretical calculations accurately reproduce the
experimental results.  These results offer the intriguing
possibility of future devices using coherent phonons to interact
with the Mn spins in much the same manner that photoexcited
carriers have been used to manipulate ferromagnetism.

We thank Xiangfeng Wang for technical help.  This work was
supported by NSF (through DMR-0134058, DMR-0325474, and
INT-0221704), DARPA (through MDA972-00-1-0034), and ONR (through
N000140410657).


\end {document}